\documentclass{article}

\usepackage{PRIMEarxiv}
\usepackage[utf8]{inputenc} % allow utf-8 input
\usepackage[T1]{fontenc}    % use 8-bit T1 fonts
\usepackage{hyperref}       % hyperlinks
\usepackage{url}            % simple URL typesetting
\usepackage{listings}
\usepackage{amsmath}
\usepackage{booktabs}       % professional-quality tables
\usepackage{amsfonts}       % blackboard math symbols
\usepackage{nicefrac}       % compact symbols for 1/2, etc.
\usepackage{microtype}      % microtypography
\usepackage{ragged2e}
\usepackage{lipsum}
\usepackage{fancyhdr}       % header
\usepackage{graphicx}       % graphics
\usepackage[sorting=none, style=nature]{biblatex}
\bibliography{references}
\graphicspath{{images/}}     % organize your images and other figures under images/ folder

\title{Kramers-Wannier Duality at the Heart of Traffic}

\author{
  Goktug Islamoglu \\
  Freiburg im Breisgau\\
  \texttt{goktugislamoglu@gmail.com} \\
}

\begin{document}
\maketitle

\begin{abstract}
\justify
To model high-risk traffic densities a cellular automaton model is constructed, exhibiting Ising model-like properties. The attraction-repulsion forces between vehicles are evaluated as coupling, and the coupling function $p(1-p)=g/8$, where p is the initial distribution of cells with state 1, and g is the Moore neighbor count, which has roots $\cos^2(\pi/8)$ and $\sin^2(\pi/8)$ for $g=1$. This is achieved without the use of any trigonometric functions in the code. From the roots, the tangent polynomial $\tan^2(x) + \tan(x)$ emerges. Kramers-Wannier duality is recovered and it is conjectured that the $1/\sqrt{2}$ difference between the roots serves as a fixed point for a projection mechanism from the 2-dimensional Ising model onto a 1-dimensional Ising chain through the Gudermannian function. The sigmoid evaluated at the proposed fixed point is substituted into the derivative of the logistic coupling function, $\sigma(1/\sqrt{2})(1-\sigma(1/\sqrt{2}))$, yielding a numerical approximation to the three-dimensional Ising inverse critical coupling. Finally, the results are linked to risk densities in traffic and vehicle types, accounting for the amplification of fatal accidents.
\end{abstract}

% Keywords (separated by commas to prevent alignment tab errors)
\keywords{Traffic Model, Ising Model, Cellular Automata}

\section{Introduction}
\justify
There are a multitude of cellular automata modeling traffic flow, some of them dating back to 1980s. Possibly the earliest of them is the Rule 184. In 1987, \cite{LI1987} the first paper on this one-dimensional cellular automaton was published. Being a one-dimensional cellular automaton, Rule 184 is fit to model single-lane traffic on a highway \cite{Boccara_1998}. Another famous one is the Nagel-Schreckenberg model. Abbreviated $NaSch$, it can model jamming of traffic using adaptable speed levels \cite{Eisenblaetter_1998}, and is an analog of Rule 184 when the speed is kept at 1. Often traffic models arise from discovering interesting patterns and trying to fit these patterns into workable traffic flow behavior. The author has done extensive traffic data exploration and analysis beforehand \cite{Islamoglu_2018}, trying to answer one question: \textit{Accidents form traffic buildup, does traffic cause accidents in retrospect?} Note that by traffic, traffic jam is not meant, but rather the general flow that is in between free-flow and full traffic jam. Renown traffic researcher and physicist Boris Kerner \cite{PhysRevLett.81.3797} calls this the \textit{synchronized flow} in his three-phase traffic theory. It is estimated by law enforcement that almost 95\% of fatal crashes are due to human error in China \cite{UNDRR2025}. Similar statistics are available from all over the world. It can be assumed that there are logical actors in traffic, such as drivers, bike riders and pedestrians, who are actively avoiding crashes, aiming to reach their destination unharmed. This is why the leading accident factors are identified as human mistakes. But what if there was \textit{another disturbance in the traffic flow, elevating the risk of humans making these mistakes?} In official reports these disturbances are often invisible and gone unnoticed. The exploratory analysis suggests points of interest, such as an airport or stadium, amplifying accident risks with poor-designed roads, leading to sudden losses in synchronization \cite{2594}. Therefore it is aimed to build a traffic model that simulates these elevated risks through traffic flow behavior in two dimensions, to capture the driving motion, lane merging conflicts and synchronization between cars.      

\section{The Traffic Model}
\label{sec:headings}
\justify

The driving motion is fundamentally the act of occupying a space on the road for a specific time period. For the next time period, another space is occupied, typically the one ahead. Vehicles can occupy transversal spaces by turning right or left, or reverse their motion by driving backwards, but these are beyond the scope of this work.

For ease of modeling, each vehicle is going to occupy a single cell on a grid, and will continue its motion forwards. A cell with state 1 will be assumed to contain a vehicle, and an empty cell with state 0 will be equated to an empty road space. 

\begin{figure}[h!tbp]
  \centering
  \includegraphics[scale=0.5]{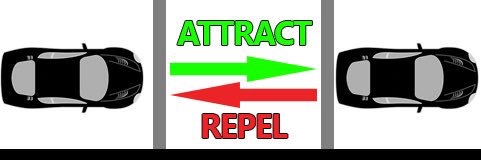}
  \caption{Attraction-repulsion forces are inherent in driving}
  \label{fig:fig1}
\end{figure}

As shown in Figure \ref{fig:fig1}, trailing vehicles are targeting to occupy the space of the car in front in the following time frame. This is for the model's sake called \textit{attraction}. The accepted sentiment is that the vehicle in front will continue moving forward. In case the vehicle in front stops, or slows down drastically, there is a risk of collision with the trailing vehicle. To mitigate this risk, the trailing vehicle is obligated to leave a trailing distance behind the car in front of it. In the model, this will be called \textit{repulsion}. When both attraction and repulsion are present, it is customary to call this effect the \textit{attraction-repulsion forces.}

Attraction-repulsion forces are common within the molecular models, such as the Lennard-Jones Potential, which has been used in vehicular modelling \cite{QU2022}. However, this work will focus on the coupling aspect, or the synchronization between vehicles, and the coupling of neighbors and next-nearest neighbors will be investigated in the context of the Ising model. 

\section{Cellular Automaton and Ising Model}
\label{sec:Ising}
\justify

Rule 184, a traffic flow model cellular automaton as mentioned before, is deemed as a kinetic Ising model for the swapping between the states of cell 1 and cell 0 \cite{PhysRevA.38.4271}. This way, continuously evolving driving motion can be obtained. In the context of this paper, however, a quenching model is obtained where the evolution of the state is ground to a halt. There are reasons for this behavior: the primary reason is the modeling of attraction-repulsion forces and lane competition. The other reason is to find out the maximum density of a traffic flow under positional competition. To determine this maximal state through comparison with other states, a frozen \textit{snapshot} is necessary. 

To obtain this snapshot, the evolution of the cellular automaton has to be inhibited. For this purpose, the author proposes the inverse update rule: A cellular automaton on a two dimensional grid can be considered to be updated based on a rule that depends on its
neighbors in all four directions of the grid \cite{Wolfram2002} for the von Neumann neighborhood case. Instead, the code will feature a cell that updates its neighbors only when its state is $0$ or $1$, such as \cite{Code201801}:

\begin{lstlisting}
for x in range(L):
    for y in range(L):
        g = number_of_Moore_neighbors(x, y)
        if c[x, y] == 0:
            nc[x, y] = 0 if g <= 6 else 1
            array0.append(c[x, y])
        elif c[x, y] == 1:
            array1.append(c[x, y])
            ...
\end{lstlisting}

where $LxL$ is the size of the lattice. 

\begin{figure}
  \centering
  \includegraphics[width=1\textwidth]{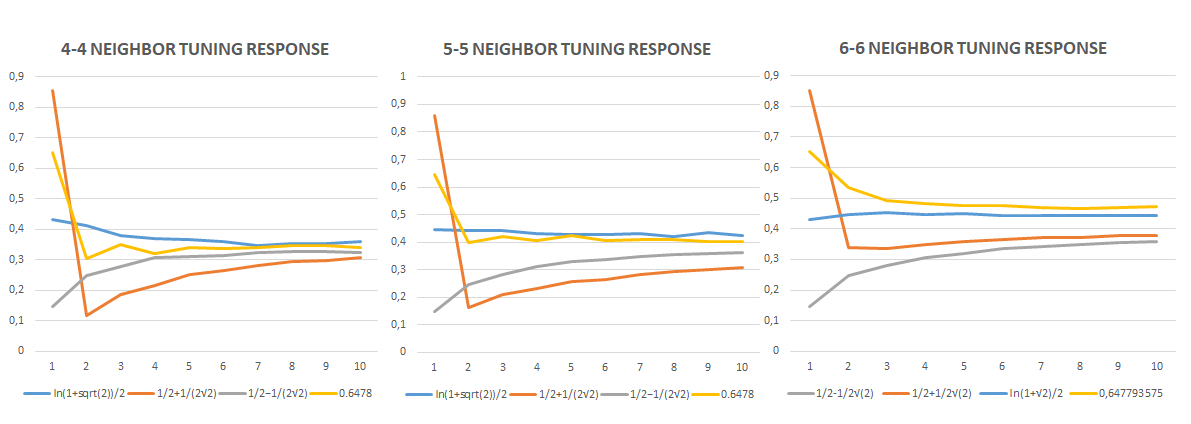}
  \caption{Tuning for 4 and 5 neighbors has inverse Ising critical temperature as the highest state 1 cell count. For 6 neighbors however, there is another maximum cell count with state 1, corresponding to magnetization.}
  \label{fig:fig2}
\end{figure}

\textit{PyCX 0.3 Realtime Visualization Template} is used for constructing the automaton simulation \cite{PyCX}. The automaton is at first initialized and cells are given state $1$ with probability $p$ \cite{Code201802}:

\begin{lstlisting}
def init():
    global c, nc, slope0, slope1, delta, o
    c = zeros([L, L])
    for x in range(L):
        for y in range(L):
            c[x, y] = 1 if random() < p else 0
    nc = zeros([L, L])
\end{lstlisting}

There is a search function applied to the top, bottom, right and left of the cell's Moore neighborhood. The top neighborhood's search function is given below as an example \cite{Code201803}:

\begin{lstlisting}
def number_of_upper_neighbors(x, y):
    upper_count = 0
    for dx in range(-1, 2):
        upper_count += c[(x + dx) % L, (y + 1) % L]
        # print upper_count
    return upper_count
\end{lstlisting}

These search functions are used as an upper limit for the driving motion. For instance, if there is one cell with state $1$ in the top neighborhood, the top-middle neighbor cell's state is updated to $1$. Same applies for bottom-middle \cite{Code201804}:

\begin{lstlisting}
m = number_of_upper_neighbors(x, y)
if m == 1:
    nc[x, (y + 1) % L] = 1
n = number_of_lower_neighbors(x, y)
if n == 1:
    nc[x, (y - 1) % L] = 1
\end{lstlisting}

The lateral driving motion is achieved by adding cells with state $1$ to the right and subtracting cells with state $1$ from the left neighborhood \cite{Code201805}:

\begin{lstlisting}
k = number_of_right_neighbors(x, y)
if k == 0 and (m <= 1 or n <= 1):
    nc[(x + 1) % L, (y + z) % L] = 1

l = number_of_left_neighbors(x, y)
if l == 1 and (m > 1 or n > 1):
    nc[(x - 1) % L, (y + z) % L] = 0
\end{lstlisting}

Then, the Moore neighborhoods and von Neumann neighborhoods are tuned against a certain threshold. When this threshold is lower than 6, such as 4 and 5, the highest cell count with state 1 is the inverse Ising temperature. When the threshold is set to 6 for both von Neumann and Moore neighborhoods, the system magnetizes to a new maximum as shown in Figure \ref{fig:fig2}.

The $p(1-p)$ function is the coupling which is central to the traffic modeling.

\begin{equation}
\sigma(x) = \frac{1}{1+e^{-x}}
\end{equation}

whose derivative is:

\begin{equation}
\sigma'(x) = \sigma(x)(1-\sigma(x))
\end{equation}

substituting $p = \sigma(x)$:

\begin{equation}
p' = p(1-p)    
\end{equation}

where p is the initial distribution of the cells with state 1. 

\begin{figure}[h!tbp]
  \centering
  \includegraphics{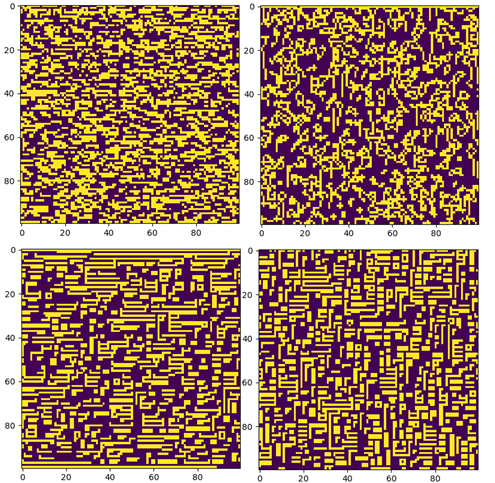}
  \caption{Clockwise: Neighborhood update without coupling; Coupling without neighborhood update; upper and lower first-order transitions respectively}
  \label{fig:fig3}
\end{figure}

Inside the code, the initial probability is coupled with its inverse and then set equal to the g divided by 8, where g is the Moore neighborhood count for every cell. There are 8 Moore neighbors of a cell, and 8 is thus selected as the denominator. 
\begin{equation}
p(1-p)=\frac{g}{8}
\end{equation}

This equality is given in the code in the following relation \cite{Code201806}:

\begin{lstlisting}
if g / 8 > (1-p) * p: 
    nc[(x + 1) % L, y] = 1
elif g / 8 < (1-p) * p:
    nc[(x - 1) % L, y] = 1
else:
    nc[x, y] = 1
\end{lstlisting}

It is important to recognize this coupling equation as central to the model. In Figure \ref{fig:fig1}, the model without any coupling is given at top left, with only driving motion. This model has no criticality. The model with only coupling and no driving motion doesn't have an evolution, only tiny vertex-like structures. Cell's evolution begins with the roots of the coupling equation, given below, shown at the bottom of the Figure \ref{fig:fig1}. \\ 

The driving motion which evolves the system specifically works if there is at most one neighbor in the system. If there aren't any neighbors, $p^2 - p = 0$ gives $p = 0$ and $p = 1$ as roots. Therefore it is impossible to discern criticality at a discrete system with binary values of $0$ and $1$. If there are two neighbors, the fraction simplifies to $p(1-p)=1/4$ which is the maximum of the $p(1-p)$ function. Its roots $1/2$ are Mean-Field inverse critical Temperature of 1D Ising Model, where the Ising model is known to don't exhibit any phase transitions \cite{MeanField2018}. Note that the $p(1-p)$ function has a maximum at $1/4$ and if it is set in that configuration, the automaton loses the first-order phase transition points, as the entire range between $[0, 1]$ becomes critical. Any number beyond $2$ results in imaginary roots. At the critical threshold $g=1$,

\begin{equation}
p(1-p)=\frac18
\end{equation}

which gives

\begin{equation}
p^2-p+\frac18=0.
\end{equation}

The roots are

\begin{equation}
p=\frac12\pm\frac{1}{2\sqrt2}.
\end{equation}

\begin{figure}[h!tbp]
  \centering
  \includegraphics[scale=0.5]{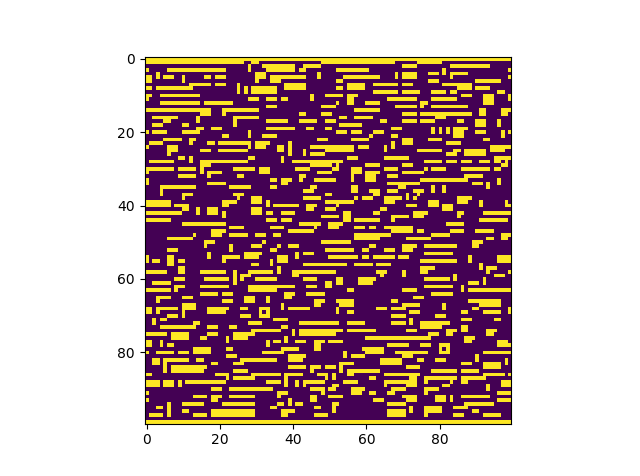}
  \includegraphics[scale=0.5]{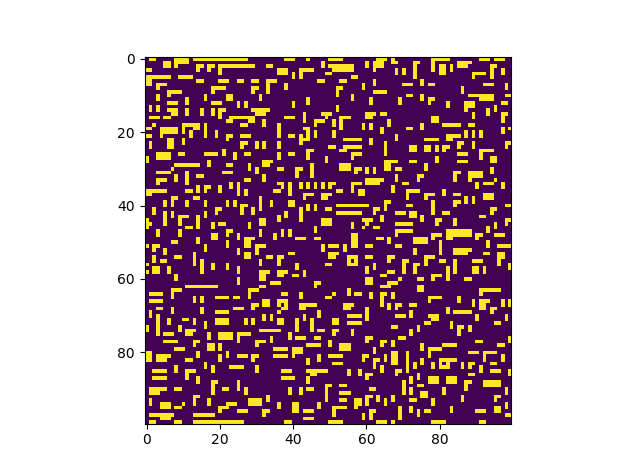}
  \caption{Uncoupled probability values do not show criticality.}
  \label{fig:fig4}
\end{figure}

\begin{figure}[h!tbp]
  \centering
  \includegraphics[scale=0.2]{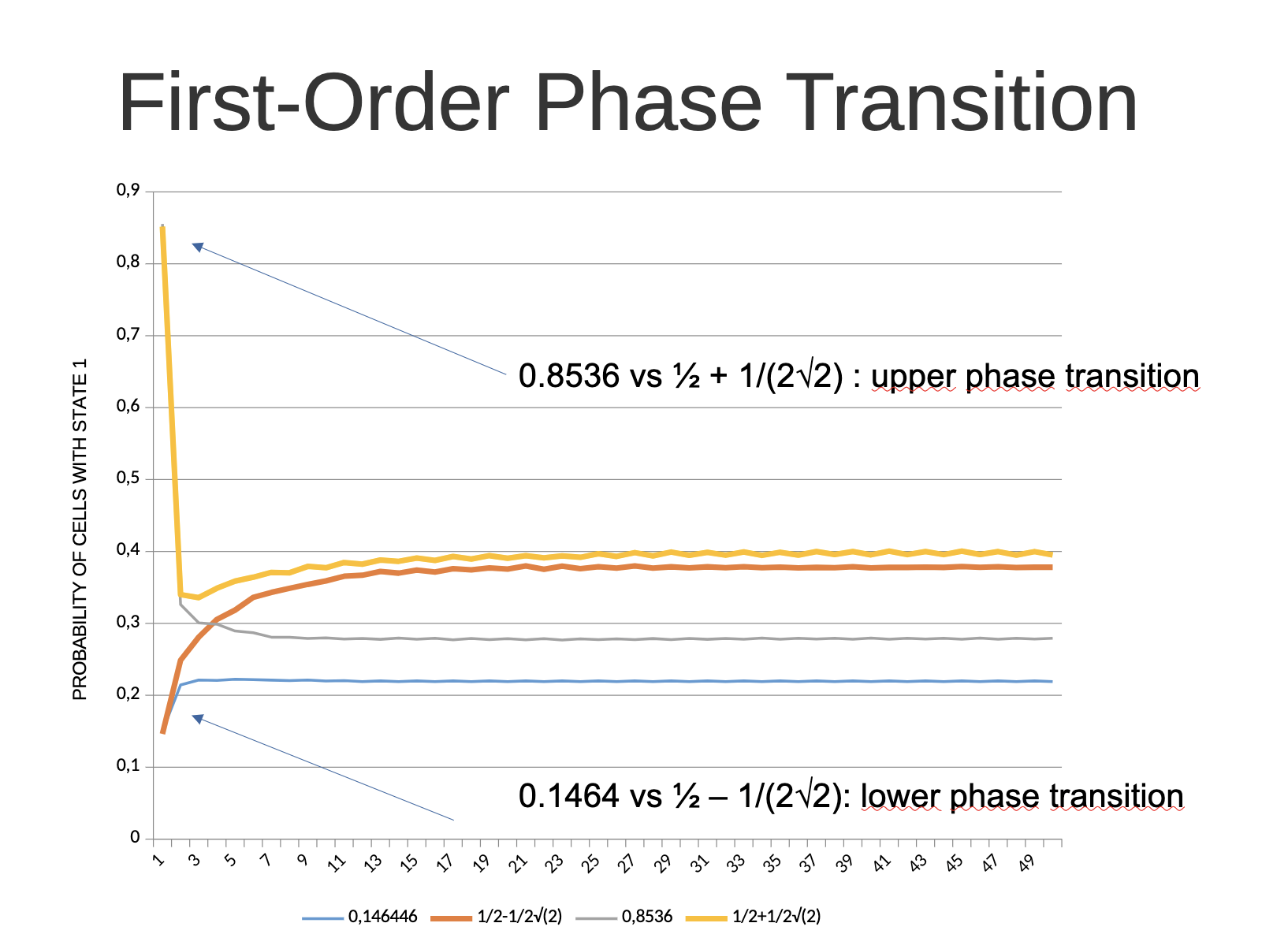}
  \caption{Two first-order phase transitions coexist in the model.}
  \label{fig:fig5}
\end{figure}

These correspond exactly to

\begin{equation}
p=\cos^2\left(\frac{\pi}{8}\right),
\qquad
p=\sin^2\left(\frac{\pi}{8}\right).
\end{equation}

where these roots exhibit first-order phase transition.
Outside of the critical boundary the initial probabilities become uncritical and lose cell counts with state 1. These figures are shown in Figure \ref{fig:fig4} and the first-order phase transitions are given in the probability plot in Figure \ref{fig:fig5}. 

\section{Outputting Tangent Equation without any Trigonometric Code}
\label{sec:tan}
\justify
The interesting output of this model is the tangent equation. Without any trigonometry in the cellular automaton, the general tangent equation is derived from the roots of the model:

\begin{equation}
\cos^2(\pi/8) = 1/2 + 1/2\sqrt{2}
\end{equation}
\begin{equation}
\sin^2(\pi/8) = 1/2 - 1/2\sqrt{2}
\end{equation} 
\begin{equation}
\cos^2(\pi/8) - \sin^2(\pi/8) = 1/\sqrt{2} = 2\sin(\pi/8)\cos(\pi/8)
\end{equation} 

\justify General Expression:
\begin{equation}
a\sin^2(x) - b\cos^2(x) = \sin(x)\cos(x)
\end{equation} 

\justify which is simplified to:
\begin{equation}
b\cot^2(x) + \cot(x) - a = 0
\end{equation}

\begin{figure}[h!tbp]
  \centering
  \includegraphics[scale=0.5]{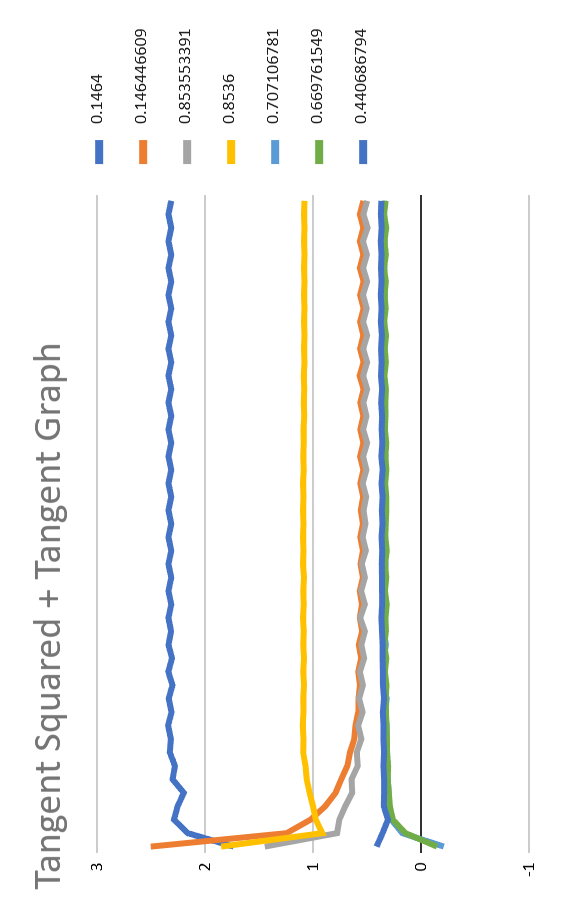}  
  \caption{Coupled and uncoupled values are arms of the tangent graph.}
  \label{fig:fig6}
\end{figure}

\justify when these values are substituted into the code using transformations, we obtain this tangent graph generated by the Ising model as seen in Figure \ref{fig:fig6}.

From the tangent outputting equation \cite{Code201807}

\begin{lstlisting}
print (ratio3 / float(ratio1 * ratio1) + (1 / float(ratio1)) - ratio2)
\end{lstlisting}

the following relations are obtained:

\begin{gather}
ratio = \frac{count0}{count1}
\\[4pt]
ratio0 = \frac{j}{i}
\\[4pt]
ratio1 = \frac{ratio0}{ratio}
\\[4pt]
\frac{ratio3}{ratio1^{2}} + \frac{1}{ratio1} - ratio2
\end{gather}

where $count0$ and $count1$ are counts of cells with state $0$ and $1$ respectively. $j$ is the sum of Moore neighborhoods in the system and $i$ is the total cell count, which is $count0 + count1$. It is deduced that $1/ratio1$ approximates to $\tan(x)$ from empirical tests. $ratio2$ and $ratio3$ are differentials of $ratio$ and $ratio0$ respectively.

$j = 8 * count1$ exact relation is obtained from simulation runs. When inserted into $ratio1$:

\begin{gather}
\frac{1}{ratio1}
=
\frac{count0\,(count0+count1)}{8\,count1^{2}}
=
\tan(x)
\\[4pt]
\end{gather}

With the evolved end-state of cells with state 1 being $count1$, the above equation produces the output of the tangent equation that was given in Figure \ref{fig:fig6}. Further simplifications are planned in future work.

\section{Kramers-Wannier Duality and Gudermannian Function}
\label{sec:Kramers-Wannier}
\justify
The exact two-dimensional Ising critical coupling,

\[
K_c=\frac12\ln(1+\sqrt2),
\]

is instead defined in hyperbolic coordinates through the Kramers--Wannier self-duality condition. Self-duality requires the high- and low-temperature expansions of the partition function to match under $K \leftrightarrow K^*$, where
\[
e^{-2K^*} = \tanh(K).
\]
At the self-dual point $K=K^*=K_c$, this fixes $K_c$ through
\[
\sinh(2K_c)=1, 
\]
equivalently written as 
\[
\sinh^2(2K_c)=1.
\] 
Solving $\sinh(2K_c)=1$ for the coupling gives $2K_c=\ln(1+\sqrt2)$, so that 
\[
e^{2K_c}=1+\sqrt2,
\qquad
e^{-2K_c}=\frac{1}{1+\sqrt2}=\sqrt2-1,
\]
the last equality following from rationalizing the denominator, $\tfrac{1}{1+\sqrt2}\cdot\tfrac{\sqrt2-1}{\sqrt2-1}=\sqrt2-1$. This is the Kramers--Wannier identity connecting the critical coupling to the golden-ratio-adjacent constant $\sqrt2-1$, and it is also equal to the automaton's recovered angle through
\begin{equation}
\tan(\pi/8) = \sqrt2 - 1 = e^{-2K_c}.
\end{equation}
The identity

\[
e^{-2K_c}=\sqrt2-1
\]

is therefore identical to

\[
\tan\!\left(\frac{\pi}{8}\right),
\]

showing that the Kramers--Wannier critical coupling is encoded by the
same half-angle tangent that appears in the automaton's algebraic
transformation.

The natural mathematical object connecting circular angles to hyperbolic coordinates is the Gudermannian function. It provides a direct correspondence between the Euclidean angle recovered by the automaton and the hyperbolic parameter governing the Ising critical point.
The Gudermannian function is defined by
\[
\mathrm{gd}(x)=2\arctan(e^x)-\frac{\pi}{2},
\]
whose inverse is
\[
\mathrm{gd}^{-1}(\phi) = \ln\!\left( \tan\left(\frac{\phi}{2}+\frac{\pi}{4}\right) \right).
\]
Evaluating the inverse Gudermannian at the self-dual angle $\phi = \pi/4$ gives
\[
\mathrm{gd}^{-1}\!\left(\frac{\pi}{4}\right)
=
\ln\!\left(\tan\frac{3\pi}{8}\right)
=
\ln(1+\sqrt2),
\]
which is exactly twice the square-lattice Ising inverse temperature,
\[
2K_c=\ln(1+\sqrt2).
\]
Equivalently, taking the negative and exponentiating recovers the Kramers--Wannier identity directly from the Gudermannian transition:
\[
e^{-\mathrm{gd}^{-1}(\pi/4)} = e^{-2K_c} = \sqrt2 - 1.
\]
Consequently,
\[
\tanh\!\left(\ln(1+\sqrt2)\right)
=
\frac{(1+\sqrt2)^2-1}{(1+\sqrt2)^2+1}
=
\frac{1}{\sqrt2},
\]
and the Gudermannian stereographic projection half-angle identity
\[
\tan\!\left(\frac{\phi}{2}\right)
=
\tanh\!\left(\frac{\psi}{2}\right)
\]
yields, at $\psi = 2K_c = \ln(1+\sqrt2)$,
\[
\tan\!\left(\frac{\pi}{8}\right)
=
\tanh\!\left(\frac{\ln(1+\sqrt2)}{2}\right)
=
\sqrt2-1
=
e^{-2K_c}.
\]
This closes the loop: the automaton's empirically recovered angle $x\approx\pi/4$ maps, through the Gudermannian transition $\mathrm{gd}^{-1}$, onto the exact hyperbolic self-dual coupling $2K_c=\ln(1+\sqrt2)$, and the Kramers--Wannier constant $e^{-2K_c}=\sqrt2-1$ reappears identically as $\tan(\pi/8)$ — the same half-angle tangent value produced independently by the tangent equation of Section \ref{sec:tan}.\\

In Section \ref{sec:Ising}, $p'$ was equated to $p(1-p)$. $\frac{1}{\sqrt{2}} \times \left(1 - \frac{1}{\sqrt{2}}\right) = \frac{1}{2} \times \tan(\pi/8)$. This result is equal to derivative of $\sigma(x)$.   

\section{1D Projection: Max Probability of the System}
\label{sec:headings}
\justify

It was established above that the difference of two critical bounds of the cellular automaton model were:

\begin{equation}
\cos^2\left(\frac{\pi}{8}\right) - \sin^2\left(\frac{\pi}{8}\right) = \frac{1}{\sqrt{2}}
\end{equation}

1D Ising Model's probability of finding a next spin up considering the first spin being found up is \cite{Lesson9}:

\begin{equation}
\frac{1}{2} + \frac{1}{2}\tanh(\beta j)
\end{equation}

If $\beta j = \ln(1+\sqrt{2})/2$, then for 2D square lattice Ising inverse critical temperature:

\begin{equation}
\frac{1}{2} + \frac{1}{2}\tanh\left(\frac{\ln(1+\sqrt{2})}{2}\right) = \frac{1}{\sqrt{2}}
\end{equation}

Thus, it shows that the $\frac{1}{\sqrt{2}}$ is an elemental part of the projection of 2D Ising model onto 1D Ising chain. \\

This is followed by McCoy's statement of the nearest neighbor correlations for the critical temperature as \cite{mccoy1994connectionstatisticalmechanicsquantum}:

\begin{equation}
\langle \sigma_{0,0}\sigma_{0,1} \rangle = \frac{2}{\pi}\coth(2K_c)\mathrm{gd}(2K_c)
\end{equation}

for critical temperature $T_c$. Substituting critical Gudermannian value:

\begin{equation}
\boxed{
\langle \sigma_{0,0}\sigma_{0,1} \rangle = \frac{2}{\pi}\sqrt{2}\frac{\pi}{4} = 1/\sqrt{2}
}
\end{equation}

$\frac{1}{\sqrt{2}}$ plays a major role in the system, as not only the anchor point of the tangent equation derived above, but also the half-angle relation that is critical to deriving the Kramers-Wannier Duality. It is the irrational reciprocal that is both the hyperbolic tangent and the sigmoid of $\ln(1+\sqrt{2})$ and the only number to do both.

\begin{equation}
\tanh(\ln(1+\sqrt{2}))
= \sigma(\ln(1+\sqrt{2}))
= \frac{1}{\sqrt{2}}
\end{equation}

Applying the sigmoid function and calculating the probability of the next spin of a 1D Ising Model to $\frac{1}{\sqrt{2}}$ implies the same result:

\begin{equation}
\sigma\left(\frac{1}{\sqrt{2}}\right) = \frac{1}{2} + \frac{1}{2}\tanh\left(\frac{1}{2\sqrt{2}}\right) = 0.66976\dots    
\end{equation}

\begin{figure}[h!tbp]
  \centering
  \includegraphics[scale=0.7]{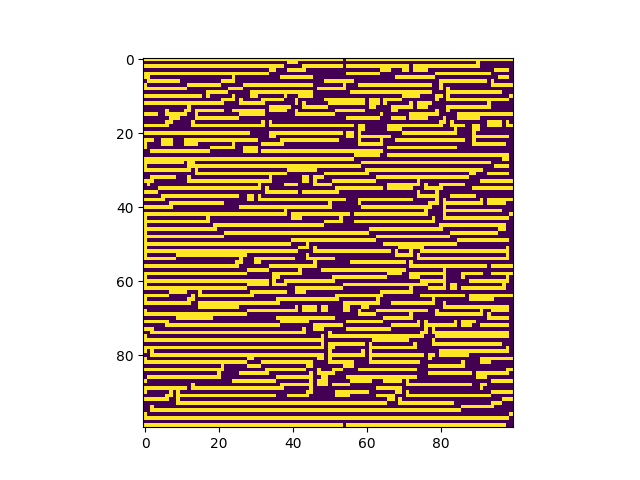}
  \caption{Highest cell count is the anisotropic extension unilaterally.}
  \label{fig:fig7}
\end{figure}

This number is the maximum probability that the 1D Ising Model we have derived with the stereographic projection. Exercising the cellular automaton with this probability confirms the highest cell output with state 1. Comparing the snapshot in Figure \ref{fig:fig7} to the critical roots in Figure \ref{fig:fig3} shows the anisotropic behavior.\\

\begin{figure}[h!tbp]
  \centering
  \includegraphics[scale=0.7]{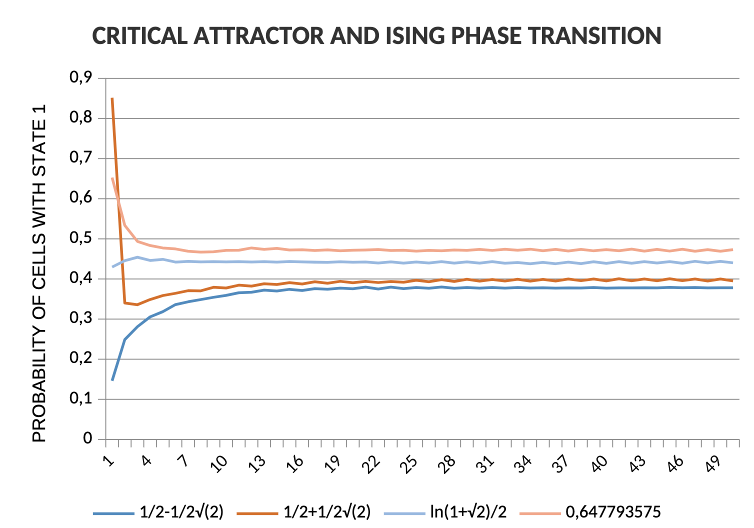}
  \caption{The model converges towards inverse Ising critical temperature.}
  \label{fig:fig8}
\end{figure}

There are two different types of phase transitions present in the system. Both first-order, as seen in Figure \ref{fig:fig5}, and second-order, presented in Figure \ref{fig:fig8} coexist in the model. Second-order phase-transition occurs at the inverse square-lattice Ising critical temperature, the Onsager result. 

Substituting the probability of the next spin to the derivative equation:\\ 
\begin{equation}
\sigma'=\sigma(1-\sigma)
\end{equation}\\
\begin{equation}
\sigma(1/\sqrt{2})(1-\sigma(1/\sqrt{2})) \approx 0.22118\dots  
\end{equation}
\\ which is only in $0.21\%$ error of $K_c^{(3D)}\approx 0.22165$, 3D inverse critical temperature of the Ising model found by the Monte Carlo method \cite{ALTalapov_1996}. The result suggests that as the 1D Ising chain extends, it facilitates a bond between the 2D lattice and 3D structures. 

\section{Traffic Hypothesis: Traffic occurs after accidents. Do Accidents Happen due to traffic flow?}
\label{sec:headings}
\justify

Paradoxes are no strangers to traffic. Just like the Jevon's Paradox predicts the increase in automobile use which comes with fuel efficiency and thus, keeps the fuel consumption at a high rate \cite{Sketch2018}, building more lanes on a roadway does not necessarily decrease traffic, but on the contrary increases it due to the perception of the users of the large roadway. This is called Induced Demand \cite{HERS-ST2002}, and the whole phenomenon is named Rebound Theory.  

Risk homoeostatis theory is another example. Peltzman argued that despite the expectations of seat belt introduction into roads to lower fatalities by 20 percent, the legislation had no effect \cite{76fd207d-24cd-363d-9999-fed29fb32f1e}. The effect in play is the perception of lower risk roads and driving riskier in response, meeting a new risk equilibrium. On the contrary, changing of left-hand driving to right-hand driving in Sweden had an astounding effect as it dramatically reduced fatalities and injuries until people got used to driving that way \cite{BMJ-2002}, which normalized the fatality and injury rate.

In Istanbul, crashes happened at the highest on middays rather than mornings and evenings where the road usage is higher, according to the Insurance Information and Monitoring Center's 2013 statistics \cite{TSB-2013}. The highest accident rates happen at a former highway - turned into roadway, D100 and the location with the highest accidents is Zeytinburnu, followed by Küçükçekmece \cite{Korkmaz2024}. 

There is a methodological error in sampling of traffic accident data. The fatalities and injuries statistics are stored at the police department, while the accidents resulting in material damage are stored at the insurance companies. The small accidents should be regarded as the heralds of fatalities and injuries, but this is not where the misconceptions end.

Crashes are seen as the result of human error. While essentially a human error has to be involved, since majority of the drivers are not motivated to crash, the factors leading humans to there errors are overlooked. The synchronization of the drivers are well-documented. The factors breaking this synchronization and causing humans to leap to errors must be investigated.

Thus, the question of are accidents caused by traffic would lead to the answer: sudden magnetization in the synchronized traffic creates a high-accident risk area, between $0.58\%$ and $0.75\%$ road density, as opposed to traffic jams and free driving, where accidents are less likely to occur. $0.66976\%$ is the critical risk density as calculated in the above section: it suggests a scale-free criticality. 

\begin{figure}[h!tbp]
  \centering
  \includegraphics[scale=0.4]{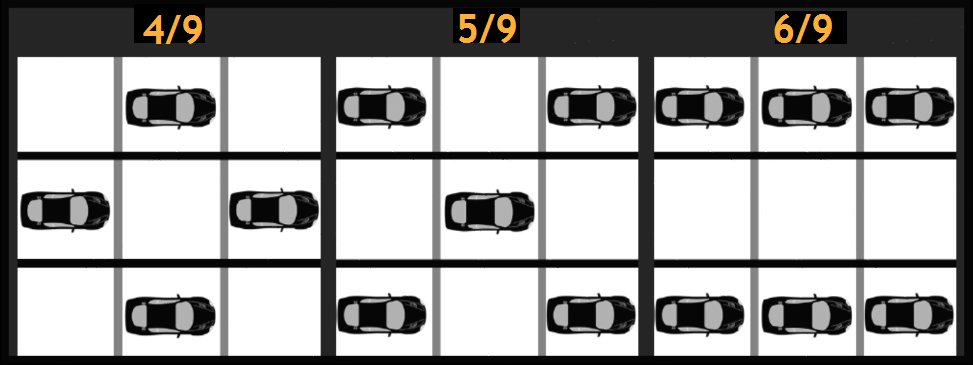} 
  \caption{6/9 is the magnetizing ratio where the only symmetry-preserving distribution is the anisotropy.}
  \label{fig:fig9}
\end{figure}

The emergence of $0.66976$ is not a coindicence, as the model's von Neumann and Moore neighborhoods are tuned to 6-6 neighbors respectively, and the magnetization of a traffic in a 3x3 lattice occurs at the $6/9$ density ratio, shown in the Figure \ref{fig:fig9}.

\section{Conclusion}
\label{sec:conc}
\justify

The emergence of a cellular automaton model that generates a tangent function, while matching the elementary identities of Ising model, is on the surface looking like it has nothing to do with traffic modeling. Traffic models are often analogies of fluid flow. The author investigates the synchronization inherent in the traffic instead \cite{Greenshields2011}. The attraction-repulsion forces in driving are magnetic in nature, and the Ising model's emergence from the roots of the critical probabilities shows the role the magnetization of the traffic flow plays in amplifying traffic accidents.

There are first-order and second-order phase transitions coexisting in the model. Without any trigonometric functions inside the code, the automaton outputs a tangent equation, which stem from the first-order transition critical roots. The second-order phase transition is the Ising criticality, as the automaton evolves towards the inverse Ising critical temperature, the Onsager result, which is derived from the Kramers-Wannier equation.

Critical accident density, which corresponds to the $0.66976$ value found in the magnetization probability, is an approximation to the 3x3 lattice of $6/9$ density, $0.66$. The first-order roots correspond to free-flow and traffic jam states, in which it is very unlikely to experience traffic accidents. Hence, they are coupled. The magnetic states and the dimensionality behind the model will be investigated in future work. 

%Bibliography
\nocite{*}
\printbibliography

@book{Wolfram2002,
  Author = {Wolfram, Stephen},
  Title = {A New Kind of Science},
  Year = {2002},
  Publisher = {Wolfram Media},
  ISBN = {1579550088},
  URL = {https://www.wolframscience.com},
  Language = {English}
    }

@article{LI1987,
  author = {Wentian Li},
  year = {1987},
  pages = {107-130},
  title = {Power Spectra of Regular Languages
and Cellular Automata},
  volume = {1},
  journal = {Complex Systems},
}

@proceedings{Greenshields2011,
    title = {75 Years of the Fundamental Diagram for Traffic Flow Theory},
    year =  {2011},
    publisher = {Greenshields Symposium},
    journal = {Transportation Research Circular}
}

@article{Boccara_1998,
   title={Cellular automaton rules conserving the number of active sites},
   volume={31},
   ISSN={1361-6447},
   url={http://dx.doi.org/10.1088/0305-4470/31/28/014},
   DOI={10.1088/0305-4470/31/28/014},
   number={28},
   journal={Journal of Physics A: Mathematical and General},
   publisher={IOP Publishing},
   author={Boccara, Nino and Fuks, Henryk},
   year={1998},
   month=July, pages={6007–6018} }

@book{2594,
	author = {McShane, William R and Roess, Roger P.},
	title = {Traffic engineering /},
	publisher = {Prentice Hall,},
	year = {c1990.},
	address = {USA,}
}

@article{Eisenblaetter_1998,
   title={Jamming transition in a cellular automaton model for traffic flow},
   volume={57},
   ISSN={1095-3787},
   url={http://dx.doi.org/10.1103/PhysRevE.57.1309},
   DOI={10.1103/physreve.57.1309},
   number={2},
   journal={Physical Review E},
   publisher={American Physical Society (APS)},
   author={Eisenblätter, B. and Santen, L. and Schadschneider, A. and Schreckenberg, M.},
   year={1998},
   month=Feb, pages={1309–1314} }

@conference{Islamoglu_2018,
    author = {Goktug Islamoglu},
    title = {Traffic Synchronization Protocol: Ising Model Phase Transitions in Hyperbolic Lattices of Accidents},
    year = {INFORMS 2018}
}

@booklet{Lesson9,
    author = {Leonard Susskind},
    title = {Lesson 9: The Ising Model} 
}

@techreport{TSB-2013,
    title = {Sigorta Bilgi ve Gözetim Merkezi: Kaza Sayısı Azalıyor},
    institution = {Insurance Association of Türkiye (TSB)},
    year = {2013} 
}

@booklet{MeanField2018,
    author = {Franz Utermohlen},
    title = {Mean Field Theory Solution of the Ising Model},
    url = {https://cpb-us-w2.wpmucdn.com/u.osu.edu/dist/3/67057/files/2018/09/Ising_model_MFT-25b1klj.pdf}
}

@misc{mccoy1994connectionstatisticalmechanicsquantum,
      title={The connection between statistical mechanics and quantum field theory}, 
      author={Barry M. McCoy},
      year={1994},
      eprint={hep-th/9403084},
      archivePrefix={arXiv},
      primaryClass={hep-th},
      url={https://arxiv.org/abs/hep-th/9403084}, 
}

@article{ALTalapov_1996,
doi = {10.1088/0305-4470/29/17/042},
url = {https://doi.org/10.1088/0305-4470/29/17/042},
year = {1996},
month = {sep},
publisher = {},
volume = {29},
number = {17},
pages = {5727},
author = {A L Talapov and H W J Blöte},
title = {The magnetization of the 3D Ising model},
journal = {Journal of Physics A: Mathematical and General}
}

@article{PhysRevA.38.4271,
  title = {Universality classes for deterministic surface growth},
  author = {Krug, J. and Spohn, H.},
  journal = {Phys. Rev. A},
  volume = {38},
  issue = {8},
  pages = {4271--4283},
  numpages = {0},
  year = {1988},
  month = {Oct},
  publisher = {American Physical Society},
  doi = {10.1103/PhysRevA.38.4271},
  url = {https://link.aps.org/doi/10.1103/PhysRevA.38.4271}
}

@article{Korkmaz2024,
    author = {Hüseyin Korkmaz},
    title = {Araç Kaza Verilerine Dayalı Trafik Kaza Süresinin Tahmini},
    year = {2024}
}

@article{76fd207d-24cd-363d-9999-fed29fb32f1e,
 ISSN = {00223808, 1537534X},
 URL = {http://www.jstor.org/stable/1830396},
 author = {Sam Peltzman},
 journal = {Journal of Political Economy},
 number = {4},
 pages = {677--725},
 publisher = {University of Chicago Press},
 title = {The Effects of Automobile Safety Regulation},
 urldate = {2026-07-11},
 volume = {83},
 year = {1975}
}

@article{BMJ-2002,
    author = {Gerald J S Wilde, Leon S Robertson and I Barry Pless},
    title = {Does risk homoeostatis theory have implications for road safety},
    journal = {BMJ},
    year = {2002}
}

@techreport{Sketch2018,
    author = {Sketchplanations},
    title = {Jevon's Paradox},
    institution = {Sketchplanations},
    year = {2018}
}

@techreport{HERS-ST2002,
    author = {Douglass B. Lee, Jr.},
    title = {Induced Demand and Elasticity},
    institution = {U.S. Department of Transportation},
    year = {2002}
}

@techreport{UNDRR2025,
    author = {United Nations Office for Disaster Risk Reduction (UNDRR), \& International Science Council (ISC)},
    title = {Road Traffic Accident}, 
    url = {https://www.undrr.org/terms/hips/TL0405},
    year = {2025}
}

@article{PhysRevLett.81.3797,
  title = {Experimental Features of Self-Organization in Traffic Flow},
  author = {Kerner, B. S.},
  journal = {Phys. Rev. Lett.},
  volume = {81},
  issue = {17},
  pages = {3797--3800},
  numpages = {0},
  year = {1998},
  month = {Oct},
  publisher = {American Physical Society},
  doi = {10.1103/PhysRevLett.81.3797},
  url = {https://link.aps.org/doi/10.1103/PhysRevLett.81.3797}
}

@article{QU2022,
  author = {Qu, Dayi and Zhao, Zixu and Hu, Chunyan and Wang, Tao and Song, Hui},
  year = {2022},
  month = {01},
  pages = {1-11},
  title = {Car-Following Dynamics, Characteristics, and Model Based on Interaction Potential Function},
  volume = {2022},
  journal = {Journal of Advanced Transportation},
  doi = {10.1155/2022/5274056}
}

@misc{PyCX,
  author = {Hiroki Sayama},
  title = {PyCX},
  year = {2025},
  url = {https://github.com/hsayama/PyCX},
}

@misc{Code201801,
  author = {Goktug Islamoglu},
  title = {Equation Automata},
  year = {2018},
  url = {https://github.com/goektug/Equation-Automata/},
  note = {{Code Lines:} 141-148},
}

@misc{Code201802,
  author = {Goktug Islamoglu},
  title = {Equation Automata},
  year = {2018},
  url = {https://github.com/goektug/Equation-Automata/},
  note = {{Code Lines:} 45-51},
}

@misc{Code201803,
  author = {Goktug Islamoglu},
  title = {Equation Automata},
  year = {2018},
  url = {https://github.com/goektug/Equation-Automata/},
  note = {{Code Lines:} 67-72},
}

@misc{Code201804,
  author = {Goktug Islamoglu},
  title = {Equation Automata},
  year = {2018},
  url = {https://github.com/goektug/Equation-Automata/},
  note = {{Code Lines:} 151-156},
}

@misc{Code201805,
  author = {Goktug Islamoglu},
  title = {Equation Automata},
  year = {2018},
  url = {https://github.com/goektug/Equation-Automata/},
  note = {{Code Lines:} 158-164},
}

@misc{Code201806,
  author = {Goktug Islamoglu},
  title = {Equation Automata},
  year = {2018},
  url = {https://github.com/goektug/Equation-Automata/},
  note = {{Code Lines:} 170-175},
}

@misc{Code201807,
  author = {Goktug Islamoglu},
  title = {Equation Automata},
  year = {2018},
  url = {https://github.com/goektug/Equation-Automata/},
  note = {{Code Lines:} 208},
}
\end{document}